%Paper: hep-th/9512108
%From: Vipul Periwal <vipul@puhep1.Princeton.EDU>
%Date: Thu, 14 Dec 1995 13:51:54 -0500

\magnification=\magstep1
\hfuzz=4truept
\nopagenumbers
\baselineskip=18truept
\font\tif=cmr10 scaled \magstep3

\rightline{PUPT-1567}
\vfil
\centerline{\tif Halpern-Huang directions in effective scalar field
theory}
\vfil
\centerline{{\rm Vipul
Periwal}\footnote{${}^\dagger$}{vipul@puhep1.princeton.edu}}
\bigskip
\centerline{Department of Physics}\centerline{Princeton University
}\centerline{Princeton, New Jersey 08544-0708}
\vfil
\par\noindent Halpern and Huang recently showed that there are relevant
directions in the space of interactions at the Gaussian fixed point.
I show that their result can be derived from Polchinski's form of
the Wilson renormalization group.  The derivation shows that the
existence of these directions is independent of the cutoff function
used.
\medskip
%\centerline{PACS: 11.10.Gh, 12.15.Cc, 64.60.Ah}
%\vskip 1 truein %
\vfil\eject

\def\D{{\hbox{D}}}
\def\d{{\hbox{d}}}

\def\part{\partial}
\def\wil{1}
\def\wh{3}
\def\pol{4}
In the Wilsonian[\wil] effective action approach to field theory, one
works with an action that has a finite short distance cutoff.  The
idea is that if one changes this cutoff, one can find appropriate
changes in the couplings of the action, so that physics much below the
cutoff scale is invariant.  This approach involves the presence of
arbitrary interaction terms in the action, including those that in
usual renormalizable quantum field theories would be regarded as
non-renormalizable.

Halpern and Huang[2] studied the space of all possible non-derivative
interaction terms about the Gaussian fixed point in $d$ dimensions,
and found that there are non-polynomial interaction terms that
are `asymptotically free', {\it i.e.} as the short distance cutoff is
taken to zero, these interactions also tend to zero.  This is in
contrast to polynomial potentials, which exhibit the opposite
behaviour.  This result is quite surprising, and may well have interesting
physical implications, as mentioned by Halpern and Huang[2].

The Wegner-Houghton[3] implementation of the Wilson scheme was used in
Ref.~[2], so some care had to be taken to avoid pathologies associated
with the use of a sharp cutoff.  In the present note, I want to show that
the result of Ref.~[2] can be derived very simply from Polchinski's
formulation[4] of the Wilson renormalization group.  Questions of
universality, {\it i.e.} independence from the manner of regularization,
can also be addressed easily in this approach.  The recursion relation
derived in Ref.~[2] is replaced here with a differential equation.

This note is organized as follows:  I recall Polchinski's version of
the Wilson renormalization group equation.  I reduce this equation
to the case needed for an analysis analogous to that of Ref.~[2].
I show that this equation has the same solutions as the solutions
found by other means in Ref.~[2].  Next, I point out
the physics behind these formal manipulations.
\def\D{{\hbox{D}}}

\def\Lam{a}

\def\d{{\hbox{d}}}

\def\part{\partial}

\def\cZ{{\cal Z}}
\def\cN{{\cal N}}
\def\parL{a{\partial\over{\partial a}}}

Start from a regulated partition function,
$$\cZ[J]\equiv\int \D\phi\ {\cN(a)}\
\exp\left(-S_a[\phi]-\int J(p)\phi(-p) \right),$$
where ${\cN(a)} $ is a normalization factor given below.
Divide $S_a=\int {1\over 2}
 \phi(x)\phi(y)a^{-2}K^{-1}((x-y)^2/a^2)$ $+ S_{a,int}.$ $a^{2}K(x^2/a^2)$
is the position-space representation of
a cutoff Green function that differs from the inverse of the
Laplacian only for large momenta.
Let
${\cN(a)} \equiv \det{}^{-{1\over2}}(K).$
We want $\parL \cZ[J]=0,$ for appropriate $J,$ corresponding to long-wavelength
correlation functions.  Writing
$$\eqalign{\cZ[J] =
\exp\left(-S_{a,int}\left[-{\delta\over{\delta J}}\right]\right)
\int \D\phi {\cN(a)}&\exp\left(-\int {1\over 2}
\phi(x)\phi(y)a^{-2}K^{-1}\left((x-y)^2/a^2
\right)\right)\cr
&\times \exp\left(-\int J(x)\phi(x) \right),\cr}$$
we find
$$\cZ[J] = \exp\left(-S_{a,int}
\left[-{\delta\over{\delta J}}\right]\right)
\exp\left(\int {1\over 2} J(x)J(y) a^2K\left((x-y)^2/a^2 \right)\right).$$
This rewriting of $\cZ[J]$ is essentially algebraic.

\def\darL#1{\Lam{{\d {#1}}\over{\d\Lam}}}
Now acting with $\parL{}$ on $\cZ[J],$ and performing some (functional)
integrations-by-parts, we find
$${\darL{S_{a,int}}} = {1\over 2}
\int\left[ {{\delta (S_{a,int}+\int J\phi)}\over\delta\phi(x)}
{{\delta (S_{a,int}+\int J\phi)}\over\delta\phi(y)}-
{{\delta^2 S_{a,int}}\over\delta\phi(x)\delta\phi(y)}\right]
{{\parL{a^2K\left((x-y)^2/a^2\right)}}}.$$
ensures that {\it all} correlation functions are left invariant.
Two comments are in order here: (1) The integrations-by-parts, in this case,
do not give any boundary terms, subject to the assumed validity of our
algebraic treatment of $\cZ[J];$ (2) For $J$ such that the Fourier
transform of $J$ vanishes for large $p$,
$J$ disappears from the above equation.

In general, $S_{a,int}$ must include arbitrary interaction terms, including
derivative interactions.
We are interested at present in examining the rg behaviour
of various interaction terms in the vicinity of the Gaussian fixed point.
In other words, we are looking at vectors in the tangent space to the
space of all interactions, at the Gaussian fixed point, so we
focus our attention on only a certain subspace of tangent vectors.
Were we to try to follow the rg trajectory away from the fixed point, we
would have to include all possible terms, as they show up in evolving
$S_{a,int}.$
If we consider $S_{a,int}= g \int a^{-d} U_a(\phi a^{d-2/2}),$
where $U_a$ is some function, for small $g$ we need only keep
$${\darL{S_{a,int}}} = -{1\over 2}\int
{{\delta^2 S_{a,int}}\over\delta\phi(x)\delta\phi(y)}
{{\parL{a^2K\left((x-y)^2/a^2\right)}}}.$$
In fact, since we are assuming that there are no derivatives in
$S_{a,int},$ we have just
$$ a^d a{\part\over {\part a}}a^{-d} U_a(\phi a^{d-2/2})
=-{1\over 2}{\part^2\over \part\phi^2}U_a(\phi a^{d-2/2})
\times a^{d-2}{{\parL{a^2K\left(0\right)}}}.$$
Thus,
$$a{\part\over {\part a}}U_a(x) + {d-2\over 2} xU_a'(x) -d U_a(x)=
+{1\over 2}U_a''(x) \kappa.$$
Here we have defined
$$\kappa = -a^{d-2}{{\parL{a^2K\left(0\right)}}},$$
which in momentum space is just
$$\kappa = -a^d \int {d^dp\over (2\pi)^d}
{1\over a^2} a{\part\over {\part a}}a^2\tilde K(p^2a^2).  $$
We made {\it no} assumptions about the form of the cutoff function in
this derivation, and all dependence on $K$ is through $\kappa=\kappa(K).$

\def\lam{\lambda}
Now, suppose we look for functions $U_a$ that scale as functions of $a,$
{\it i.e.} functions such that
$$a{\part\over {\part a}}U_a(x) = \lambda U_a(x).$$
Then we want to solve
$$(\lambda-d)f(x;\lam) - {\kappa\over 2}f''(x;\lam) + {d-2\over 2} xf'(x;\lam)
=0.$$
It is trivial to show that even solutions of this equation are exactly those
found by Halpern and Huang, who imposed $f(0;\lam)=0.$
The only difference as compared to Ref.~[2] is that we have found $\kappa$
for a general cutoff function, where they found the volume of a $d-1$
dimensional sphere divided by $(2\pi)^d.$  In fact, for a sharp cutoff,
$\kappa$ takes exactly the value they found.
The solutions are given in terms of the Kummer function:
$$M(a,b,c) \equiv {\Gamma(b)\over\Gamma(b-a)\Gamma(a)}
\int_0^1 dt \exp(tc) t^{a-1}(1-t)^{b-a-1},$$
with
$$f(x;\lam) = \left[ M((\lam-d)/(d-2),1/2,(d-2)x^2/2\kappa)-1\right].$$
Various properties of these potentials are explained in Ref.~[2], so I will
not go into the details.

One should not suppose that perturbation theory for such interaction
potentials is simple.  The case of interest is $\lam>0,$ since for
such eigenvalues, as $a\downarrow 0,$ $U_a$ scales down.
In fact, the full potential is
$$V(\phi) \equiv ga^{\lambda-d} U_1(\phi a^{d-2/2}).$$
Thus the `anomalous' dependence on $a$ is as $a^\lam,$ which is
much stronger than the logarithmic scale dependence one finds for
dimensionless couplings in asymptotically free renormalizable
field theories.
$V(\phi)$ is to be treated in its entirety as
a perturbation, in contrast to usual field theory where one rescales
the interaction monomials so that higher order monomials are
multiplied by higher powers of the coupling constant.
What this implies is that even at the lowest order in perturbation
theory one has to sum over an infinite set of graphs.  For any $n$-point
connected correlation function, at first order in perturbation theory
the contributions are from graphs with
a single vertex, but there are vertices of arbitrarily high degree, implying
that most of the contractions are of tadpole type.  A moment's reflection
will make it evident that this is precisely what is encoded in the
differential equation that determines these potentials.
The scaling behaviour is dependent on the existence of such an infinite
series of terms.

It is important to note, as well, that one cannot draw immmediate
conclusions about the scaling behaviour of low energy correlation
functions when their arguments are scaled from the behaviour of the
coupling parameter in the action.  In fact, low energy correlation
functions are independent of $a$ by construction.  One must compute to
second order in the perturbation before one finds a nontrivial $\beta$
function with the same interpretation as the $\beta$ functions of
renormalizable field theories---as mentioned above, the scaling found
above comes from tadpole type graphs, and these have {\it no}
dependence on external momenta at all.

One may wonder if the term quadratic in $S_{a,int}$ in the rg equation
would change the scaling behaviour found in Ref.~[2].
The term quadratic in $S_{a,int}$ could contribute to the
evolution of the potential, but in fact it does not if the cutoff function
satisfies
$$\lim_{p\downarrow0} {\part\over{\part a}}a^2\tilde K(p^2a^2)
=0.$$
This is a very reasonable restriction on cutoff functions, since it merely
implies that at low momenta the propagator is independent of the
cutoff.  So the linear equation stays valid for the flow of the potential
terms, except for the fact that the field needs to be rescaled at second
order, and derivative interactions in $S_{a,int}$ can also contribute to
the flow of the potential terms.

The effect of field rescaling is approximately of the form
$$f(x;\lam) = g \left[ M((\lam-d)/(d-2+2\eta),
1/2,(d-2+2\eta)x^2/2\kappa)-1\right],$$
except that $\eta=O(g^2)$ is not trivial to determine, and indeed
depends on higher moments of the cutoff propagator.  This is how
the evolution of the kinetic term
produces changes in the flow of the effective potential.
The result of Halpern and Huang[2] is clearly unaffected by such
nonlinear terms, since their calculation was expressly to determine the
relevant directions in the space of perturbations of a scalar
effective field theory.  It would be extremely interesting to study the
evolution of such relevant directions numerically.

This work was supported in part by NSF Grant No. PHY90-21984.
\hfuzz=2pt
\bigskip
\centerline{References}
\bigskip
\item{\wil.} K. Wilson, in {\it Collective properties of physical
systems}, (B. Lundqvist and S. Lundqvist, eds.),
Academic Press, New York, 1973
K.G. Wilson and J.B. Kogut, {\sl Phys. Rep.} {\bf 12C}
(1974) 75; F. Wegner, in {\it Phase transitions
and critical phenomena}, vol. 6, (C. Domb and M. Green, eds.), Academic
Press, New York, 1976

\item{2.} K. Halpern and K. Huang, {\sl Phys. Rev. Lett.} {\bf 74} (1995) 3526;
MIT report, CTP 2477 (1995)

\item{\wh} F. Wegner and A. Houghton, {\sl Phys. Rev.} {\bf A8} (1972) 401;

\item{\pol}J. Polchinski, {\sl Nucl. Phys.} {\bf B231} (1984) 269

\end